# 2001 QR$_{322}$ – an update on Neptune's first unstable Trojan companion

Jonti Horner[1,2], Patryk Sofia Lykawka[3]

[1] Computational Engineering and Science Research Centre, University of Southern Queensland, Toowoomba, Queensland 4350, Australia
[2] Australian Centre for Astrobiology, UNSW Australia, Sydney, New South Wales 2052, Australia
[3] Astronomy Group, School of Interdisciplinary Social and Human Sciences, Kinki University Shinkamikosaka 228-3, Higashiosaka-shi, Osaka, 577-0813, Japan

**Summary:** The Neptune Trojans are the most recent addition to the panoply of Solar system small body populations. The orbit of the first discovered member, 2001 QR$_{322}$, was investigated shortly after its discovery, based on early observations of the object, and it was found to be dynamically stable on timescales comparable to the age of the Solar system.

As further observations were obtained of the object over the following years, the best-fit solution for its orbit changed. We therefore carried out a new study of 2001 QR$_{322}$'s orbit in 2010, finding that it lay on the boundary between dynamically stable and unstable regions in Neptune's Trojan cloud, and concluding that further observations were needed to determine the true stability of the object's orbit.

Here we follow up on that earlier work, and present the preliminary results of a dynamical study using an updated fit to 2001 QR$_{322}$'s orbit. Despite the improved precision with which the orbit of 2001 QR$_{322}$ is known, we find that the best-fit solution remains balanced on a knife-edge, lying between the same regions of stability and instability noted in our earlier work. In the future, we intend to carry out new observations that should hopefully refine the orbit to an extent that its true nature can finally be disentangled.

## Introduction

In 2001, astronomers carrying out the Deep Ecliptic Survey ([36][37]) announced the discovery of five new trans-Neptunian objects, including one given the designation 2001 QR$_{322}$. In the discovery circular [1], the proposed orbit for that object would have made it a Plutino (e.g. [2][3]), trapped in Neptune's 2:3 mean-motion resonance. Indeed, the circular states: "*The assumed perihelic Neptune 2:3-resonance orbit for 2001 QR322 keeps the object more than 16 AU from Neptune over a 14 000-year period*".

As a Plutino, 2001 QR$_{322}$ would not have been particularly remarkable – well over a hundred have been found to date. However, it was soon realised that 2001 QR$_{322}$ was significantly more interesting. Follow-up observations in November and December 2002 led to major revisions in the best solution for the objects orbit, and in January 2003, it was announced that 2001 QR$_{322}$ was, in fact, a Neptune Trojan – the first to be discovered [4]. In that circular, it was noted that, "*[a]ccording to the above orbit, the 1:1 Neptune librator remains more than 20 AU from Neptune over a 14 000-year period. E. Chiang has confirmed the object's status as the first known "Neptune Trojan" by integrating the orbit over $10^9$ years*".

Following up on the discovery, two independent studies examined the orbital stability of the newly detected Neptunian Trojan. [5] employed a Frequency Map Analysis to produce maps of the diffusion rate of hypothetical Uranian and Neptunian Trojans. They noted that 2001 QR$_{322}$ lay close to the border of a stable region for low inclination Neptune Trojans, but found its orbit to be highly stable. They performed numerical integrations of a population of test particles based on the best-fit orbit of 2001 QR$_{322}$, and found that only 10% escaped from the Trojan cloud over the lifetime of the Solar system, inferring that the object must be primordial, rather than a recent capture.

This result was strongly supported by the work of [6], who found that whilst the orbital evolution of 2001 QR$_{322}$ within the Neptunian Trojan cloud was chaotic, perturbed by the $v_{18}$ secular resonance, this was insufficient to cause the object to escape from the Trojan cloud. In their simulations, the authors found that "*[t]he probability of escape to a non-Trojan orbit in our simulations was low, and only occurred for orbits starting near the low-probability edge of the orbital element distribution (largest values of initial semimajor axis and small eccentricity)*". The stable nature of 2001 QR$_{322}$, it seemed, was confirmed, and was thereafter assumed by studies considering the formation and evolution of the Neptunian Trojan population (e.g. [7][8]).

In 2009, we began a project to examine whether it was possible to use the distribution of the Neptune Trojan population to constrain the nature of the giant planet's outward migration. It was clear from the orbits of the Plutinos that Neptune must have migrated over a considerable distance (e.g. [9][10]), but the newly discovered Neptune Trojans offered the opportunity to test the models of planetary migration to see which, if any, could reproduce the observed distribution, and predict the range of orbits over which future Trojans would be found (e.g. [11][12][13][14]).

In the course of that work, we noticed that the best-fit orbital solution for 2001 QR$_{322}$ had changed significantly over the years since the work of [5] and [6], as a result of new observations. With a longer observational arc available, the orbit of 2001 QR$_{322}$ had become better constrained, with orbital elements that had shifted significantly from the original studies. We decided to see whether the new orbit remained dynamically stable.

To do this, we carried out detailed *n*-body dynamical simulations of the new orbit ([15]), and found that the new solution was balanced precariously between regions of dynamical stability and instability. In our simulations, test particles located at a semi-major axis greater than 30.30 au were found to be significantly less stable than those at smaller semi-major axes. The nominal best-fit solution for the semi-major axis of 2001 QR$_{322}$ at the time of that work lay at 30.3023 au, right at the very inner edge of the unstable region.

In our simulations, half of the test particles considered were removed from the Solar system within just 590 Myr – a period far shorter than the system's age, but not so short that it did not remain feasible that 2001 QR$_{322}$ had formed in the Neptune Trojan cloud, or had been captured during that planet's migration, in the Solar system's youth. Given the sharp delineation between regions with greatly differing stability, our results highlighted the critical importance of further follow-up observations of 2001 QR$_{322}$, to better constrain the objects orbit and to help resolve on which side of the dynamical divide it actually lies.

In this work we present the preliminary results of a fresh study of 2001 $QR_{322}$'s dynamical stability, based on an updated orbital solution published on the Asteroids Dynamics Site (AstDyS; [16]) website in January 2014. In the next section we describe the set-up of our new simulations, before presenting our preliminary results. We then conclude with a discussion of those results, and of the future work we plan to undertake.

## The Simulations

In order to study the stability of the orbit of 2001 $QR_{322}$, we took the best-fit orbital solution from the Asteroids Dynamic Site, AstDyS ([16]) on 24[th] January 2014. The full solution, with uncertainties, is given in Table 1. For comparison, we also present the orbital elements used in our previous work, given in red italicised text, as obtained on 26[th] January 2009 from the same website. Whilst the uncertainty on the semi-major axis, $a$, has increased slightly, the uncertainties of the other elements have dropped by a factor of ~2. We note, also, that the value of some elements has changed by significantly more than their original uncertainties[1]. Taken in concert, this highlights the critical importance of follow-up observations for Solar system objects – by extending the observational arc over which an object has been followed, its orbit can be greatly refined.

| **Element** | **Value** | **1-σ uncertainty**[2] |
|---|---|---|
| ***a* (au)** | 30.2848 | 0.009292 |
|  | *30.3023* | *0.008813* |
| ***e*** | 0.027298 | 0.0001605 |
|  | *0.031121* | *0.0003059* |
| ***i (°)*** | 1.322 | 0.0005654 |
|  | *1.323* | *0.0009417* |
| ***Ω (°)*** | 151.599 | 0.01484 |
|  | *151.628* | *0.02328* |
| ***ω (°)*** | 163.415 | 0.4789 |
|  | *160.73* | *0.8316* |
| ***M (°)*** | 66.336 | 0.4753 |
|  | *57.883* | *0.7818* |
| **Epoch (JD)** | 2456600 |  |
|  | *2454800* |  |

*Table 1: The best-fit orbital elements, and their associated 1-σ uncertainties, for 2001 $QR_{322}$. The values given in black are those used in the current work, and were obtained from [16] on 24[th] January 2014. The values given in red italics are those used in our previous work ([15]), obtained from [16] on 26[th] January 2009, and are included here to show how the best-fit solution has changed as a result of new observations being made. Here, a is the semi-major axis, e the eccentricity, i the inclination, Ω the longitude of the ascending node, ω the longitude of perihelion and M the mean anomaly of 2001 $QR_{322}$ at the epoch given in the final row.*

---

[1] The value of the semi-major axis, for example, has changed by 0.0175 au, a shift of twice the stated uncertainty in the 2009 values. The changes in $M$ and $\omega$ are even more striking – with their values shifting my many times the stated 2009 uncertainties.

[2] We note, here, that the values presented in this table, and used in our work, are the precise values as taken from the AstDyS system. It is readily apparent that the uncertainties in a given value (column 3) stretch to significantly more significant figures than given for the 'best fit' value (column 2). Whilst this is not ideal, we felt it best to exactly reproduce the data as taken from the AstDyS website, to ensure reproducibility of our results.

To study the long-term stability of the orbit of 2001 QR$_{322}$, we used the Hybrid integrator within the *n*-body dynamics package MERCURY ([17]) to follow the orbital evolution of an ensemble of massless test particles under the gravitational influence of the four giant planets for a period of four billion years. This technique (following the long term evolution of many objects using MERCURY) has proved highly successful in determining the dynamical stability of objects in our own Solar system (e.g. [15][18][19]). It has also been used to probe questions in astrobiology (e.g. [20][21][22][23]) and exoplanetary science (e.g. [24][25][26]). It allows the chaotic orbital evolution of the objects in question to be quantified in a probabilistic sense, as well as enabling the creation of stability maps that allow the behaviour as a function of initial orbital elements to be examined.

As in our earlier work, we created an ensemble of clones of 2001 QR$_{322}$ for our integrations, centred on the nominal best-fit orbit. The clones were spread across the full ±3σ uncertainty range for each of the six orbital elements considered. In our earlier studies of both Solar system objects (e.g. [27][28]) and exoplanetary systems (e.g. [25][26]), we have found that the orbital semi-major axis and eccentricity of objects typically play by far the most important role in determining their long-term stability – a fact that was clearly demonstrated in our earlier studies of 2001 QR$_{322}$ ([15]). For that reason we constructed our ensemble to maximise our resolution in these orbital elements. We tested a total of 51 unique values of semi-major axis, distributed evenly across the full ±3σ uncertainty range. At each of these semi-major axes, we tested 51 unique eccentricities, again evenly spread across the full ±3σ uncertainty range in that element. At each of these locations in *a-e* space, we tested five discrete values of inclination, and three unique values of the longitudes of perihelion and ascending node, and three mean anomalies, each spread evenly over the full ±3σ uncertainty range[3]. In total, this produced a test sample of 351,135 test particles (51 x 51 x 5 x 3 x 3 x 3).

The orbits of these test particles were then integrated forward in time under the gravitational influence of the four giant planets, Jupiter, Saturn, Uranus and Neptune, with an integration time-step of 120 days. Objects were removed from the simulations when they collided with one of the giant planets, fell into the central body, or were ejected to a barycentric distance of 1000 au, as in our previous studies of the Neptunian and Jovian Trojans ([15][18][27][28]). When objects were removed in this manner, the time at which the ejection or collision occurred was recorded.

## Preliminary Results

Of the 351,135 clones of 2001 QR$_{322}$ integrated in this work, 99,949 survived for the full four billion years of the integrations (a 28.46% survival rate). During the first few million years of the simulations, no ejections or collisions occurred – with the first test particle to be removed coming after 3.66 million years. After this slow start, the decay began to proceed apace, with fully half the test particles being removed in the first 600 million years. By recording the time at which each test particle was removed, we are able to create a dynamical map that illustrates how the stability of 2001 QR$_{322}$ depends on its initial orbit, as can be seen in Figure 1, below. Equally, one can plot the decay of the number of test particles as a function of time, as can be seen in Figure 2.

---

[3] We therefore tested the best-fit inclination value along with four others, located -3σ, -1.5σ, +1.5σ and +3σ from that value. For longitude of perihelion, ascending node, and mean anomaly, we therefore tested the best-fit values along with those +3σ and -3σ away.

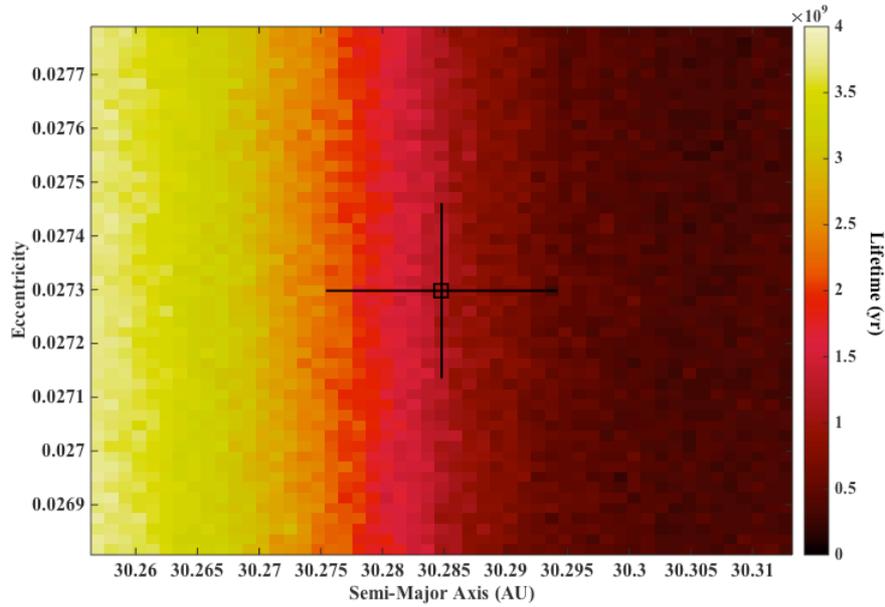

*Figure 1: The mean dynamical lifetime of the orbit of 2001 QR$_{322}$ as a function of its initial semi-major axis and eccentricity. The nominal best-fit orbit lies at the centre of the plot, within the small hollow box, and the horizontal and vertical lines that radiate from it represent the 1-σ uncertainties in semi-major axis and eccentricity. The lifetime plotted at each a-e location is the mean of 135 discrete simulations. It is clear that the best-fit orbital solution for 2001 QR$_{322}$ falls right on the boundary between a region of high stability and one that is far more chaotic.*

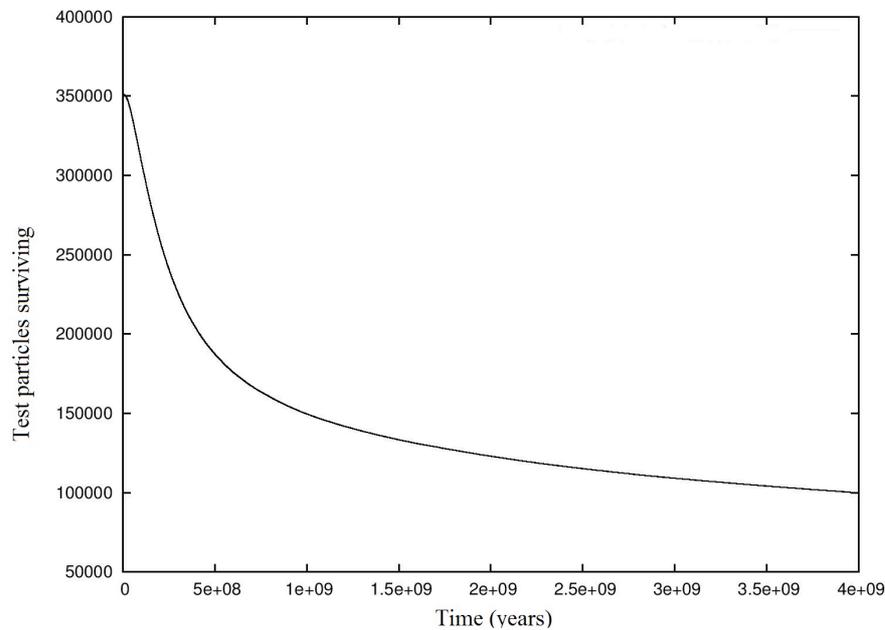

*Figure 2: The number of test particles surviving as a function of the time elapsed in our simulations. After a short period of dynamical relaxation (the first particle removed survived for 3.66 Myr), the population decays in a broadly exponential manner, although the 'half-life' of that exponential decay increases as time passes, as the objects from the least stable reaches of a-e space are removed.*

## Discussion and Conclusions

In our previous study of the stability of 2001 QR$_{322}$ ([15]), we examined the evolution of an ensemble of 19,683 test particles, cloned in a six-dimensional swarm centred on the nominal best-fit orbit available at the time. That work revealed an object balanced on a dynamical precipice – right on the boundary between stable and unstable regions.

With a more refined orbit, based on a longer observational arc, we anticipated that new simulations could answer the question of 2001 QR$_{322}$'s stability. On one hand, the best-fit solution might have moved to a more stable region of orbital element space, and we could therefore conclude that 2001 QR$_{322}$ was most likely captured as a Neptune Trojan in the final stages of planetary migration (e.g. [12]), or had formed with the planet and been carried along with it as the planet migrated to its current resting place (e.g. [11]). On the other, it was equally possible that the improved orbit for 2001 QR$_{322}$ could place it firmly in an unstable region – suggesting that it might only recently have been captured as a Neptunian Trojan (as is thought to be the case for 2004 KV$_{18}$; [28]).

In this work we expand upon our previous study of 2001 QR$_{322}$, taking advantage of the greatly increased computational capacity available to us. We simulated a population of test particles more than an order of magnitude larger than in our previous work, and followed their evolution for a factor of four times longer (4 Gyr vs. 1 Gyr). This has allowed us to increase the resolution with which we can map the dynamical stability of 2001 QR$_{322}$'s orbit, as well as allowing us to probe cases where the instability occurs on timescales longer than those considered in our earlier work.

Whilst our detailed analysis of the results is still ongoing, it is immediately apparent from examination of Figure 1 that the stability of 2001 QR$_{322}$ remains strongly dependent on the initial semi-major axis of its orbit. Despite the longer observational arc and generally smaller uncertainties in the best-fit orbit, that solution remains precariously balanced between regions of significant dynamical stability and instability. As in our earlier work, fully half of the test particles considered were removed from the Solar system within the first 600 Myr of our simulations. Indeed, despite the changes to the orbital solution, our results are remarkably similar to those we obtained in 2010.

Our simulations reveal that all *a-e* locations within the 1-σ uncertainties are unstable on timescales less than ~2 Gyr, which in turn suggests that the object is truly dynamically unstable. That said, it should be noted that instability on timescales of hundreds of millions of years, or several billion years, is not incompatible with the idea that 2001 QR$_{322}$ has been trapped in the Neptunian Trojan cloud since the cessation of planetary migration (as discussed in [15]).

To illustrate this point, let us assume, for the sake of argument, that the initial population of the Neptunian Trojan cloud was large[4], and then assume that a non-negligible subset of that population moved on orbits similar to that of 2001 QR$_{322}$. In such a scenario, it is readily apparent that a significant population of such objects could survive to the current day. For a population decaying with a dynamical 'half-life' of 600 Myr, ~1% would be expected to survive for four billion years. Furthermore, as we argued in [30], such a scenario would allow the Neptunian Trojans to act as a continual source of fresh material to the Centaur population, and from there, to the inner Solar system[5].

---

[4] This seems a reasonable assumption, given that estimates of the *current* Neptunian Trojan population have ranged as high as $10^7$ objects more than 1 km in diameter (e.g. [29]).

Given that objects escaped from the Neptunian Trojan cloud at all locations plotted in Figure 1, it is clear that even the most stable orbital solutions for 2001 QR$_{322}$ do not preclude its eventual escape to the Centaur population. We still require the orbit of 2001 QR$_{322}$ to be further constrained before we can conclusively determine its true stability.

To that end we plan to obtain follow-up observations of 2001 QR$_{322}$ in 2016, with the goal of further refining the best-fit solution to its orbit, and hopefully answering the question of its true dynamical stability, once and for all.

## Acknowledgements

JH is supported by USQ's Strategic Research Fund: the STARWINDS project. The authors thank Richard Schwarz and an anonymous reviewer for their comments and feedback on our work during the review process, which helped us to improve the clarity and flow of our paper.

## References

1. Wasserman, L. H., Chiang, E., Jordan, A. B., Ryan, E. L., Buie, M. W., Millis, R. L., Kern, S. D., Elliot, J. L., Washburn, K. E. and Marsden, B. G., " MPEC 2001-V11 : 2001 QQ322, 2001 QR322, 2001 RU143, 2001 RV143, 2001 RW143", *Minor Planet Center Electronic Circular 2001-V11*, 2001, available at http://www.minorplanetcenter.net/iau/mpec/K01/K01V11.html

2. Yu, Q. and Tremaine, S., "The Dynamics of Plutinos", *The Astronomical Journal*, 1999, 118, pp. 1873 – 1881

3. Chiang, E. I. and Jordan, A. B., "On the Plutinos and Twotinos of the Kuiper Belt", *The Astronomical Journal*, 2002, 124, pp. 3430 – 3444

4. Pittichova, J. A., Meech, K. J., Wasserman, L. H., Trilling, D. E., Millis, R. L., Buie, M. W., Kern, S. D., Clancy, K. B., Hutchison, L. E., Chiang, E. and Marsden, B. G., "MPEC 2003-A55 : 2001 QR322", *Minor Planet Center Electronic Circular 2003-A55,* 2003, available at http://www.minorplanetcenter.net/iau/mpec/K03/K03A55.html

5. Marzari, F., Tricarico, P. and Scholl, H., "The MATROS project: Stability of Uranus and Neptune Trojans. The case of 2001 QR322", *Astronomy and Astrophysics*, 2003, 410, pp. 725 – 734

6. Brasser, R., Mikkola, S., Huang, T.-Y., Wiegert, P., and Inanen, K., "Long-term evolution of the Neptune Trojan 2001 QR322", *Monthly Notices of the Royal Astronomical Society*, 2004, 347, pp. 833 – 836

---

[5] The Centaurs are the direct parent population to the short-period comets (e.g. [31][32]), and are likely themselves sourced from a variety of regions, including the Scattered Disk (e.g. [33]), the Oort Cloud (e.g. [34][35]), and the Jovian and Neptunian Trojan clouds (e.g. [30]).

7. Chiang, E. I. and Lithwick, Y., "Neptune Trojans as a Test Bed for Planet Formation", *The Astrophysical Journal*, 2005, 628, pp. 520 – 532

8. Nesvorný, D. and Vokrouhlický, D., "Chaotic Capture of Neptune Trojans", *The Astronomical Journal*, 2009, 137, pp. 5003 – 5011

9. Malhotra, R., "The origin of Pluto's peculiar orbit", *Nature*, 1993, 365, pp. 819-821

10. Hahn, J. M. and Malhotra, R., "Orbital Evolution of Planets Embedded in a Planetesimal Disk", *The Astronomical Journal*, 1999, 117, pp. 3041 – 3053

11. Lykawka, P. S., Horner, J., Jones, B. W. and Mukai, T., "Origin and dynamical evolution of Neptune Trojans – I. Formation and planetary migration", *Monthly Notices of the Royal Astronomical Society*, 2009, 398, pp. 1715 – 1729

12. Lykawka, P. S. and Horner, J., "The capture of Trojan asteroids by the giant planets during planetary migration", *Monthly Notices of the Royal Astronomical Society*, 2010, 405, pp. 1375 – 1383

13. Lykawka, P. S., Horner, J., Jones, B. W. and Mukai, T., "Formation and dynamical evolution of the Neptune Trojans – the influence of the initial Solar system architecture", *Monthly Notices of the Royal Astronomical Society*, 2010, 404, pp. 1272 - 1280

14. Lykawka, P. S., Horner, J., Jones, B. W. and Mukai, T., "Origin and dynamical evolution of Neptune Trojans – II. Long-term evolution", *Monthly Notices of the Royal Astronomical Society*, 2011, 412, pp. 537 - 550

15. Horner, J. and Lykawka, P. S., "2001 QR322: a dynamically unstable Neptune Trojan?", *Monthly Notices of the Royal Astronomical Society*, 2010, 405, pp. 49 - 56

16. The AstDyS website, http://hamilton.dm.unipi.it/astdys/ ; described in Knezevic, Z. and Milani, A., "Asteroids Dynamic Site – AstDyS", *IAU Joint Discussion 7: Space-Time Reference Systems for Future Research at IAU General Assembly - Beijing*, 2012

17. Chambers, J. E., "A hybrid symplectic integrator that permits close encounters between massive bodies", *Monthly Notices of the Royal Astronomical Society*, 1999, 304, pp. 793- 799.

18. Horner, J., Müller, T. G. and Lykawka, P. S., " (1173) Anchises – thermophysical and dynamical studies of a dynamically unstable Jovian Trojan", *Monthly Notices of the Royal Astronomical Society*, 2012, 423, 2587 - 2596

19. Kiss, Cs., Szabó, Gy., Horner, J., Conn, B. C., Müller, T. G., Vilenius, E., Sárneczky, K., Kiss, L. L., Bannister, M., Bayliss, D., Pál, A., Góbi, S., Verebélyi, E., Lellouch, E., Santos-Sanz, P., Ortiz, J. L., Duffard, R. and Morales, N., "A portrait of the extreme solar system object 2012 DR$_{30}$", *Astronomy and Astrophysics*, 2013, 555, article id. A3

20. Horner, J. and Jones, B. W., "Jupiter friend or foe? I: The asteroids", *International Journal of Astrobiology*, 2008, 7, pp. 251 - 261

21. Horner, J. and Jones, B. W., "Jupiter - friend or foe? II: the Centaurs", *International Journal of Astrobiology*, 2009, 8, pp. 75 - 80

22.     Horner, J., Jones, B. W. and Chambers, J., "Jupiter - friend or foe? III: the Oort cloud comets", *International Journal of Astrobiology*, 2010, 9, pp. 1 - 10

23.     Horner, J. and Jones, B. W., "Jupiter - friend or foe? IV: The influence of orbital eccentricity and inclination", *International Journal of Astrobiology*, 2012, 11, pp. 147 - 156

24.     Marshall, J., Horner, J. and Carter, A., "Dynamical simulations of the HR8799 planetary system", *International Journal of Astrobiology*, 2010, 9, pp. 259 - 264

25.     Robertson, P., Horner, J., Wittenmyer, R. A., Endl, M., Cochran, W. D., MacQueen, P. J., Brugamyer, E. J., Simon, A. E., Barnes, S. I. and Caldwell, C., "A Second Giant Planet in 3:2 Mean-motion Resonance in the HD 204313 System", *The Astrophysical Journal*, 2012, 754, article id. 50

26.     Wittenmyer, R. A., Horner, J. and Marshall, J. P., "On the dynamical stability of the proposed planetary system orbiting NSVS 14256825", *Monthly Notices of the Royal Astronomical Society*, 2013, 431, pp. 2150 - 2154

27.     Horner, J., Lykawka, P. S., Bannister, M. T. and Francis, P., "2008 LC18: a potentially unstable Neptune Trojan", *Monthly Notices of the Royal Astronomical Society*, 2012, 422, pp. 2145 - 2151

28.     Horner, J. and Lykawka, P. S., "2004 $KV_{18}$: a visitor from the scattered disc to the Neptune Trojan population", *Monthly Notices of the Royal Astronomical Society*, 2012, 426, pp. 159 - 166

29.     Sheppard, S. S. and Trujillo, C. A., "A Thick Cloud of Neptune Trojans and Their Colours", *Science*, 2006, 5786, pp. 511 – 514

30.     Horner, J. and Lykawka, P. S., "Planetary Trojans – the main short of the short period comets?", *International Journal of Astrobiology*, 2010, 9, pp. 227 – 234

31.     Levison, H. F. and Duncan, M. J., "From the Kuiper Belt to Jupiter-Family Comets: The Spatial Distribution of Ecliptic Comets", *Icarus*, 1997, 127, pp. 13 - 32

32.     Horner, J., Evans, N. W. and Bailey, M. E., "Simulations of the population of Centaurs – I. The Bulk statistics", *Monthly Notices of the Royal Astronomical Society*, 2004, 354, pp. 798 – 810

33.     Volk, K. and Malhotra, R., "The Scattered Disk as the Source of the Jupiter Family Comets", *The Astrophysical Journal*, 2008, 687, pp. 714 – 725

34.     Brasser, R., Schwamb, M. E., Lykawka, P. S. and Gomes, R. S., "An Oort cloud origin for the high-inclination, high-perihelion Centaurs", *Monthly Notices of the Royal Astronomical Society*, 2012, 420, pp. 3396 - 3402

35.     Emel'yanenko, V. V., Asher, D. J. and Bailey, M. E., "A Model for the Common Origin of Jupiter Family and Halley Type Comets", *Earth, Moon and Planets,* 2013, 110, pp. 105-130

36. Millis, R. L., Buie, M. W., Wasserman, L. H., Elliot, J. L., Kern, S. D. and Wagner, R. M., "The Deep Ecliptic Survey: A Search for Kuiper Belt Objects and Centaurs. I. Description of Methods and Initial Results", *The Astronomical Journal*, 2002, 123, pp. 2083 – 2109

37. Elliot, J. L., Kern, S. D., Clancy, K. B., Gulbis, A. A. S., Millis, R. L., Buie, M. W., Wasserman, L. H., Chiang, E. I., Jordan, A. B., Trilling, D. E. and Meech, K . J., "The Deep Ecliptic Survey: A Search for Kuiper Belt Objects and Centaurs. II. Dynamical Classification, the Kuiper Belt Plane, and the Core Population", *The Astronomical Journal*, 2005, 129, pp. 1117 - 1162